\documentclass[10pt,twocolumn,twoside]{IEEEtran}

\IEEEoverridecommandlockouts

\usepackage{graphicx} 
\usepackage{epstopdf} 
\usepackage{cite}
\usepackage{amsmath,amssymb,amsfonts}
\usepackage{algorithmic}
\usepackage{graphicx}
\usepackage{textcomp}
\usepackage{xcolor}
\usepackage{color}

\usepackage{amssymb}
\usepackage{bbm} 
\usepackage{bm}

\usepackage{algorithm}
\usepackage{algorithmic}
\usepackage{url}
\usepackage{color}

\usepackage{graphicx} 
\usepackage{epstopdf} 

\usepackage{cite}
\usepackage{amsmath,amssymb,amsfonts}
\usepackage{algorithmic}
\usepackage{graphicx}
\usepackage{textcomp}
\usepackage{xcolor}
\usepackage{color}

\usepackage{amssymb}
\usepackage{bbm} 
\usepackage{bm}

\usepackage{algorithm}
\usepackage{algorithmic}
\usepackage{url}
\usepackage{color}

\usepackage{multirow} 

\usepackage{subfig}

\usepackage{tablefootnote}

\usepackage{caption}

\usepackage{amsmath}
\allowdisplaybreaks[1]

\usepackage{booktabs}
\usepackage{threeparttable}
\usepackage{footnote}
\usepackage{enumerate}

\def\BibTeX{{\rm B\kern-.05em{\sc i\kern-.025em b}\kern-.08em
    T\kern-.1667em\lower.7ex\hbox{E}\kern-.125emX}}
\begin{document}

\title{Spatial Privacy-aware VR streaming}
\author{\IEEEauthorblockN{Xing Wei and Chenyang Yang
}
\\
\IEEEauthorblockA{School of Electronics and Information Engineering, Beihang University, Beijing 100191, China\\ Email: \{weixing, cyyang\}@buaa.edu.cn}
}

\maketitle

\vspace{-18mm}
\begin{abstract}
    Proactive tile-based virtual reality (VR) video streaming employs the current tracking data of a user to predict future requested tiles, then renders and delivers the predicted tiles before playback. 
    Very recently, privacy protection in proactive VR video streaming starts to raise concerns. However, existing privacy protection may fail even with privacy-preserve federated learning.
    This is because when the future requested tiles can be predicted accurately, the user-behavior-related data can still be recovered from the predicted tiles.
    In this paper, we consider how to protect privacy even with accurate predictors and investigate the impact of privacy requirement on the quality of experience (QoE). To this end, we first add extra \textit{camouflaged} tile requests to the real tile requests and model the privacy requirement as the \textit{spatial degree of privacy} (sDoP). By ensuring sDoP, the real tile requests can be hidden and privacy can be protected. Then, we jointly optimize the
    durations for prediction, computing, and transmitting, aimed at maximizing the privacy-aware QoE given arbitrary predictor and configured resources. From the obtained optimal closed-form solution, we find that the impacts of sDoP on the QoE are two sides of the same coin. On the one side the increase of sDoP improves the capability of communication and computing hence improves QoE. On the other side it degrades the prediction performance hence degrades the QoE. The overall impact depends on which factor dominates the QoE. Simulation with two predictors on a real dataset verifies the analysis and shows that the overall impact of sDoP is to improve the QoE.
\end{abstract}
\vspace{-1mm}

\vspace{-4mm}
\section{Introduction}
Wireless virtual reality (VR) can provide a seamless and immersive experience to users. As the main type of VR services, 360$^\circ$ video has the following unique features.
First, 360$^\circ$ video usually has $360^{\circ}\times 180^{\circ}$ panoramic view with ultra high resolution (e.g., binocular 16K \cite{HuaWei_Cloud_VR}). Second, the range of angles of a 360$^\circ$ video that humans can see at arbitrary time is only a small portion of the full panoramic view (e.g., $110^\circ\times110^\circ$), which is called the field of view (FoV). Third, the stalls or black holes during watching 360$^\circ$ video will cause physiological discomfort, e.g., dizziness, which degrades the quality of experience (QoE) and thus should be avoided.

To stream such video with QoE guarantee, proactive VR video streaming is proposed \cite{optimizing_VR}, which divides a full panoramic view segment into small tiles in the spatial domain. Before the playback of the segment, the tiles to be most likely requested in the segment are first predicted using the user-behavior-related data in an observation window, which are then rendered and finally delivered to the user.

While proactive VR video streaming is being intensively investigated in academia and industry, most of the existing works
neglect the willingness of users. Are users willing to share their behavior-related data while watching 360$^\circ$ videos?

Recent work shows that with less than 5 minutes tracking data while watching VR videos, the random forest algorithm can correctly identify 95\% of users among all the 511 users \cite{privacy_VR_identifiability}. This indicates that behavior-related data can be used to infer personal information. With the development of technology, one may be able to dig more personal information than beyond imagination. Very recently, the first privacy requirement dataset shows that the privacy requirements of 360$^\circ$ videos among videos and users are heterogeneous, and only 41\% of the totally watched videos have no privacy requirement \cite{privacy_Xing}.

With these findings, privacy protection in VR video streaming starts to raise concerns. A privacy-preserve approach has been proposed to continuously upload one of the user-behavior-related data---eye-tracking data \cite{privacy-preserving_eye_tracking_2021}. Privacy requirement has been defined in the temporal domain and the corresponding impact on QoE has been investigated \cite{privacy_Xing}. However, proactive VR streaming still needs to predict the future requested tiles. When the prediction is accurate, the user-behavior-related data can still be recovered from the predicted tiles, and the privacy protection may fail.
Then, here comes two problems:
\textit{How to protect privacy in VR video streaming even with accurate predictors? What is the impact of privacy protection on the system?}

In this paper, we strive to answer these questions. Our contributions can be summarized as follows.

\begin{itemize}
	\item To protect privacy even with accurate predictors, we blur the real tile requests by adding extra \textit{camouflaged} tile requests. Specifically, we define the spatial degree of privacy (sDoP) as a metric related to the number of extra camouflaged requested tiles in addition to real requested tiles. Then, the input of the predictor becomes a mixture of real and camouflaged requested tiles. A larger sDoP indicates more extra camouflaged tile requests, which will degrade the prediction accuracy. By ensuring sDoP, the real tile requests can be hidden and privacy can be efficiently protected.
	\item Based on the defined privacy requirement, we optimize the durations for prediction, communication, and computing under arbitrarily given predictor as well as communication and computing resources to maximize the QoE. 
	From the obtained optimal solution, we find that the impacts of sDoP on the QoE are contradictory. One the one hand, the increase of sDoP improves the capability of communication and computing hence improves QoE. On the other hand, it degrades the prediction performance hence degrades the QoE. The overall impact depends on which factor dominates the QoE. 
	\item Simulation with two predictors on a real dataset verifies the analysis and shows that the overall impact of increasing the sDoP is to improve the QoE.
\end{itemize}

\vspace{-1.5mm}
\section{System Model}\label{section-system_model}
\vspace{-.5mm}
Consider a tile-based VR video streaming system with a multi-access edge computing (MEC) server co-located with a base station (BS) that serves $K$ users.
The MEC server equips with powerful computing units for rendering, accesses a VR video library by local caching or high-speed backhaul, thus the delay from the Internet to the MEC server can be omitted.
Each user requests 360$^{\circ}$ videos from the library according to their own interests. 
In the sequel, we consider arbitrary one request for $v$th video from $k$th user for analysis.

Each VR video consists of $L$ segments in the temporal domain, and each segment consists of $M$ tiles in the spatial domain. The playback duration of each tile equals the playback duration of a segment, denoted by $T_{\mathrm{seg}}$ \cite{optimizing_VR,survey_Hsu}.

Each user is equipped with an head mounted display (HMD), which can measure the user-behavior-related data (e.g., the head movement trace),
send the tile requests to the MEC server, and pre-buffer segments. To protect the privacy of users, the HMD is also equipped with a light-weighted computing unit for training a predictor and predicting tile requests.

\vspace{-0.4cm}
\subsection{Spatial Degree of Privacy}
\vspace{-0.4cm}
 \begin{figure}[htbp]
	\centering
	\begin{minipage}[t]{1\linewidth}
		\includegraphics[width=1\textwidth]{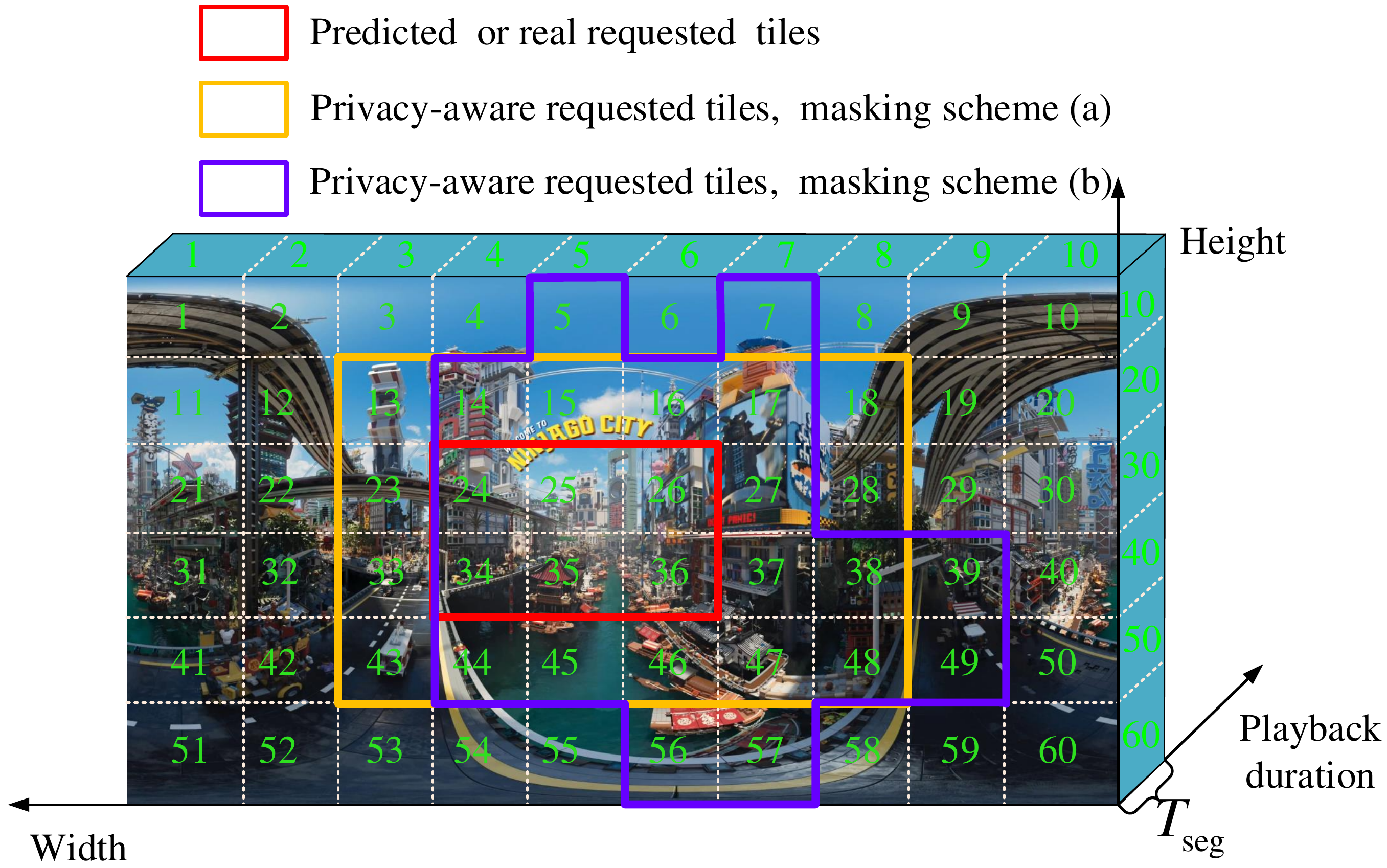}
	\end{minipage}
	\vspace{-0.2cm}
	\caption{Predicted or real requested tiles and privacy-aware requested tiles in a segment, $M=60$, $N_{\textit{fov}}=3\times2 = 6$, $N_p=6\times 4 = 24$.}
	\label{Fig:sDoP_explain}
	\vspace{-0.3cm}
\end{figure}

For training a tile request predictor or predicting the requested tiles at the MEC server by centralized learning, the requested tiles should be uploaded to the MEC server. To hide the real tile requests when a user has privacy requirement, the HMD should request a mixture of real and camouflaged requested tiles. For training at HMDs, say by federated learning, the real tile requests are only stored at the local HMD. However, when predicting at HMD, the predicted tile requests still need to be uploaded to the MEC server. When the prediction is accurate, the MEC server can still obtain the real tile requests. That is to say, \textit{even with federated learning for VR streaming, privacy will still be leaked out}.
Therefore, for the same reason of privacy protection, the HMD should also upload a mixture of predicted and camouflaged requested tiles. To reflect the privacy requirement in centralized and federated prediction when watching 360$^{\circ}$ videos, we define the \textit{spatial degree of privacy} as the ratio of extra camouflaged requested tiles except tiles in an FoV among all the extra tiles except tiles in an FoV, i.e.,
\begin{equation}\label{def:sDoP}
	\rho_{s} \triangleq \frac{N_p - N_{\textit{fov}}}{M - N_{\textit{fov}}}\in[0,100\%]
\end{equation}
where $N_p$ is the number of \textit{privacy-aware requested tiles} in a segment, which contains real or predicted requested tiles and camouflaged requested tiles, $N_{\textit{fov}}$ is the number of tiles in an FoV of a segment. Both numbers of real and predicted requested tiles are  $N_{\textit{fov}}$.  To protect privacy, $N_p\geq N_{\textit{fov}}$.
To illustrate the sDoP, we provide an example in Fig. \ref{Fig:sDoP_explain}. The number of tiles in an FoV is $N_{\textit{fov}}=6$, the predicted or real requested tiles in the FoV are No. 24-26, 34-36. To protect the privacy, varies masking schemes can be employed. For example, for masking scheme (a), the extra camouflaged requested tiles are No. 13-18, 23, 27, 28, 33, 37, 38, 43-48, and the number of privacy-aware requested tiles is $N_p=24$. Then, sDoP is $\rho_s = \frac{24 - 6}{60 - 6} = 33.3\%$. When $\rho_s = 0$, the user has no privacy requirement, the real or predicted tile requests are uploaded to the MEC server. When $\rho_s = 100\%$, the user has the most stringent privacy requirement, the HMD always requests all tiles or predicts all tiles will be requested.

When the user sets privacy requirement sDoP, the number of privacy-aware requested tiles can be obtained from \eqref{def:sDoP} as 
\begin{align}\label{def:N_p}
	N_p(\rho_s) = N_{\textit{fov}} + \rho_s (M - N_{\textit{fov}})
\end{align}

\vspace{-0.4cm}
\begin{figure}[htbp]
	\centering
	\begin{minipage}[t]{1\linewidth}
		\includegraphics[width=1\textwidth]{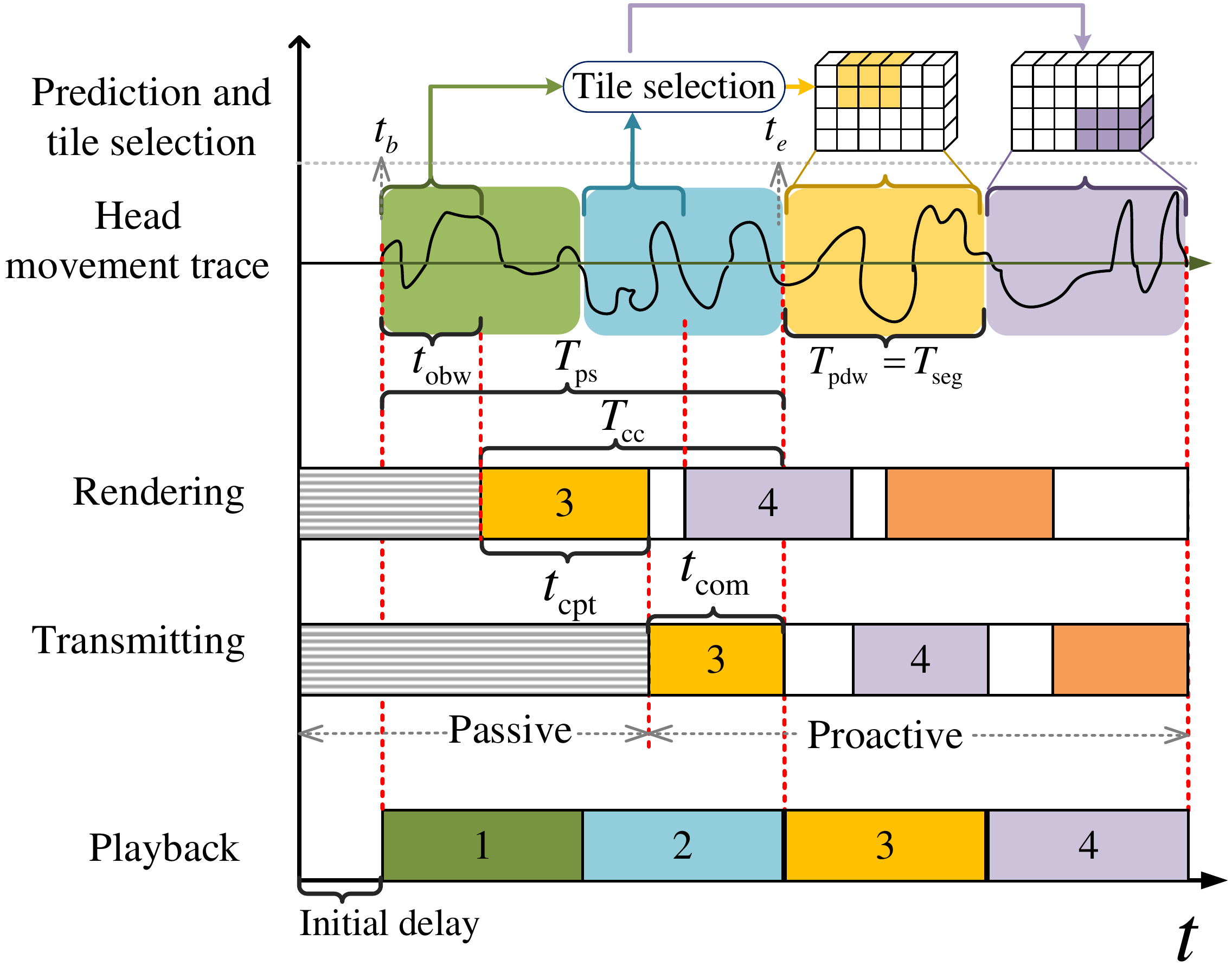}
	\end{minipage}
	\caption{Streaming the first four segments of a VR video. $t_b$ is the start time of the observation window, $t_e$ is the start time of playback of the $l_0$th segment, $l_0=3$.}
	\label{Fig:MEC4VR_pipeline}
\end{figure}
\vspace{-0.6cm}

\subsection{Streaming Procedure}

As shown in Fig. \ref{Fig:MEC4VR_pipeline}, when a user requests a VR video with sDoP $\rho_s$, the MEC server first streams the initial ($l_{0}-1$)th segments in a passive streaming mode \cite{transmission_mode-standard-update}. After an initial delay, the first segment begins to play at the time instant $t_b$, which is also the start time of the observation window. Then, proactive streaming for $l_0$th segment begins, subsequent segments are predicted, computed, and transmitted. In the sequel, we take the $l_0$th segment as an example for elaboration.

After the MEC server collects the user-behaviour-related data in an observation windows with duration $t_{\mathrm{obw}}$, the tiles to be played in the $l_0$th segment with duration $T_{\mathrm{pdw}}=T_{\mathrm{seg}}$ can be predicted. 
To avoid playback stalling, rendering and transmitting the tiles in $l_0$th segment should be finished before 
the start time of playback of $l_0$th segment, i.e., the time instant $t_e$. The duration beginning from $t_b$ and terminating at $t_e$, is the proactive streaming time for a segment $T_{\mathrm{ps}}$. We can observe that $T_{\mathrm{ps}}=(l_0 -1)T_{\mathrm{seg}}$. In Fig. \ref{Fig:MEC4VR_pipeline}, we consider predicting the third segment as an example, i.e., $l_{0}=3$, $T_{\mathrm{ps}}=2T_{\mathrm{seg}}$.

Specifically, at the end of the observation window, tile request probabilities or the fixation sequences of FoVs in the $l_0$th segment can be predicted at HMD. 
Based on the probabilities (or the fixation sequences) and the number of tiles in an FoV $N_{\textit{fov}}$, the predicted requested tiles can be obtained \cite{privacy_Xing}. Given sDoP $\rho_s$, the number of privacy-aware requested tiles $N_p(\rho_s)$ can be obtained from \eqref{def:N_p}. Then, based on the predicted requested tiles, $N_p(\rho_s)$, and a masking scheme (say scheme (b) in Fig. \ref{Fig:sDoP_explain}), the extra camouflaged requested tiles can be determined.  
The selected tiles are rendered with duration $t_{\mathrm{cpt}}$ and the sequence of FoVs can be generated, and finally the sequence of FoVs are transmitted with duration $t_{\mathrm{com}}$, which should be finished before the start time of playback for the predicted segment. 
The durations for observation, computing, and transmitting should satisfy $t_{\mathrm{obw}} + t_{\mathrm{cpt}} + t_{\mathrm{com}}=T_{\mathrm{ps}}$. The duration for communication and computing can be expressed as $t_{\mathrm{cc}}\triangleq t_{\mathrm{com}} + t_{\mathrm{cpt}}$. 



\vspace{-0.4cm}
\subsection{Computing and Transmission Model}
According to the computing model in \cite{Xing_VR_Shannon}, the number of bits that can be rendered per second, referred to as the computing rate, is $C_{\mathrm{cpt},k} \triangleq \frac{\mathcal{F}_{\mathrm{cpt},k}}{K\cdot\mu_r} (\textit{in bit/s})$,
where $\mu_r$ is the required floating-point operations (FLOPs) for rendering one bit of FoV \textit{in FLOPs/bit} \cite{Xing_VR_Shannon}.

The BS serves $K$ single-antenna users using zero-forcing beamforming with $N_t$ antennas.
The instantaneous data rate at the $i$th time slot for the $k$th user is
\begin{align*}
	C_{\mathrm{com},k}^{i}=B\log_2 \left(1+\frac{p_k^i d_k^{-\alpha}|\tilde{h}^i_k|^2}{\sigma^2} \right)
\end{align*}
where $B$ is the bandwidth, $\tilde{h}^i_k\triangleq (\mathbf{h}^i_k)^H\mathbf{w}^i_k$ is the equivalent channel gain, $p_k^i$ and $\mathbf{w}^i_k$ are respectively the transmit power and beamforming vector for the $k$th user,  $d_k$ and $\mathbf{h}^i_k\in\mathbb{C}^{N_t}$ are respectively the distance and the small scale channel vector from the BS to the $k$th user, $\alpha$ is the path-loss exponent, $\sigma^2$ is the noise power, and $(\cdot)^{H}$ denotes conjugate transpose.

We consider indoor users as in the literature, where the distances of users, $d_k$, usually change slightly \cite{survey_Hsu,NTHU_dataset} and hence are assumed fixed.
Due to the head movement and the variation of the environment,
small-scale channels are time-varying, which are assumed as remaining constant in each time slot with duration $\Delta T$ and changing independently with identical distribution among time slots. With the proactive transmission, the rendered tiles in a segment should be transmitted with duration $t_{\mathrm{com}}$. The number of bits transmitted with $t_{\mathrm{com}}$ can be expressed as $\overline{C}_{\mathrm{com},k}\cdot t_{\mathrm{com}}$, where 
\begin{align*}
	\overline{C}_{\mathrm{com},k} \triangleq \frac{1}{N_s}\sum_{i=1}^{N_s}C_{\mathrm{com},k}^{i} \cdot \Delta T
\end{align*}
is the time average transmission rate, and $N_s$ is the number of time slots in $t_{\mathrm{com}}$.
Since future channels are unknown when making the optimization, we use ensemble-average rate $\mathbbm{E}_h\{C_{\mathrm{com},k}\}$ \cite{ergodic-capacity} to approximate the time-average rate $\overline{C}_{\mathrm{com},k}$, where $\mathbbm{E}_h\{\cdot\}$ is the expectation over $h$, which can be very accurate when $N_s$ or $N_t/K$ is large \cite{Xing_VR_Shannon}. 

To ensure fairness among users in terms of QoE, the transmit power is allocated to compensate the path loss, i.e., ${p}^i_k=\frac{\beta}{d_k^{-\alpha}}$, where $\beta$ can be obtained from $\beta(\sum_{k=1}^{K}\frac{1}{d_k^{-\alpha}})={P}$ and ${P}$ is the maximal transmit power of the BS. Then, the ensemble-average transmission rate for each user is equal.

In the sequel, we consider arbitrary one user and use $C_{\mathrm{com}}$ and $C_{\mathrm{cpt}}$ to replace $\mathbbm{E}_h\{C_{\mathrm{com},k}\}$ and $C_{\mathrm{cpt},k}$ for notional simplicity. 


\vspace{-0.4cm}
\section{Problem Formulation}
\vspace{-0.2cm}

%

\subsection{Performance Metric of Tile Prediction}\label{section:DoO_def}

Average segment degree of overlap (average-DoO) has been used to measure the prediction performance for a VR video \cite{Xing_VR_Shannon}. It indicates the average overlap of the predicted tiles and the real requested tiles among all the proactively streamed segments, which is defined as
\begin{align*}
	\mathcal{D}(t_{\mathrm{obw}}) \triangleq \frac{1}{L-l_0 + 1}\sum_{l=l_0}^L\frac{\mathbf{q}_{l}^\mathsf{T}\cdot\mathbf{e}_{l}({t_{\mathrm{obw}}})  }{\|\mathbf{q}_{l}\|_1}\in[0,100\%]
\end{align*}
where $\mathbf{q}_{l}\triangleq [q_{l,1},...,q_{l,M}]^\mathsf{T}$ denotes the ground truth of the tile requests for the $l$th segment with $q_{l,m}\in\{0,1\}$, $\mathbf{e}_{l}({t_{\mathrm{obw}}})\triangleq [e_{l,1}({t_{\mathrm{obw}}}),...,e_{l,M}({t_{\mathrm{obw}}})]^\mathsf{T}$ denotes the predicted tile requests for the segment with $e_{l,m}({t_{\mathrm{obw}}})\in\{0,1\}$,
$(\cdot)^\mathsf{T}$ denotes transpose of a vector, and $\|\cdot\|_1$ denotes the $\ell_1$ norm of a vector. When the $m$th tile in the $l$th segment is truly requested, $q_{l,m}=1$, otherwise $q_{l,m}=0$. When the tile is predicted to be requested, $e_{l,m}({t_{\mathrm{obw}}})=1$, otherwise it is zero. We consider $\|\mathbf{e}_{l}\left(t_{\mathrm{obw}}\right)\|_1 = N_{\textit{fov}}$.
A larger value of average-DoO indicates a better prediction.


As the verified Assumption 1 in \cite{Xing_VR_Shannon} states, a predictor can be more accurate with a longer observation window. Therefore, average-DoO is a monotonically increasing function of $t_{\mathrm{obw}}$. 
\vspace{-0.4cm}
\subsection{Communication and Computing Capability as well as  Resources Rate}\label{section:s_com,s_cpt}
The capability of communication and computing (CC) can be used to measure the capability of streaming tiles. It is the ratio of tiles in a segment that can be rendered and transmitted with assigned transmission and computing rates and corresponding durations, i.e.,
\begin{align*}
	C_{\mathrm{cc}}(t_{\mathrm{com}}, t_{\mathrm{cpt}}) \triangleq \left. \min\left\{\frac{C_{\mathrm{com}}t_{\mathrm{com}}}{s_{\mathrm{com}}^{}}, \frac{C_{\mathrm{cpt}}t_{\mathrm{cpt}}}{s_{\mathrm{cpt}}^{}}, M\right\} \middle/ M \right.
\end{align*}
where $\min\left\{\frac{C_{\mathrm{com}}t_{\mathrm{com}}}{s_{\mathrm{com}}^{}}, \frac{C_{\mathrm{cpt}}t_{\mathrm{cpt}}}{s_{\mathrm{cpt}}^{}}, M\right\}$ is the number of tiles that can be computed and transmitted, $s_{\mathrm{com}} = {px}_w \cdot{px}_h \cdot b \cdot r_f \cdot T_{\mathrm{seg}}/\gamma_{c}$ \cite{HuaWei_Cloud_VR} is the number of bits in each tile for transmission, $s_{\mathrm{cpt}} = {px}_w \cdot{px}_h \cdot b \cdot r_f \cdot T_{\mathrm{seg}}$ is the number of bits in a tile for rendering, ${px}_w$ and ${px}_h$ are the pixels in wide and high of a tile, $b$ is the number of bits per pixel relevant to color depth \cite{HuaWei_Cloud_VR}, $r_f$ is the frame rate, and $\gamma_c$ is the compression ratio.

To reflect the capability of streaming tiles in unit time, we further define the \textit{resources rate} as
\begin{align}
	R_{\mathrm{cc}}\triangleq \frac{C_{\mathrm{cc}}(t_{\mathrm{com}}, t_{\mathrm{cpt}})}{t_{\mathrm{cc}}}\label{def:R_cc},
\end{align}

\subsection{Metric of Privacy-aware Quality of Experience}
For proactive tile-based streaming without privacy requirement, the QoE can be measured by the percentage of the correctly streamed tiles among all the real requested tiles \cite{Xing_VR_Shannon}.
When considering the spatial degree of privacy, we should also consider whether the sDoP can be satisfied. This is because if the QoE is only captured by the percentage of the correctly predicted tiles, then, the MEC server can still obtain the accurate location of real requested tiles from the feedback of QoE from the HMD.
Therefore, the privacy-aware QoE should consist of two parts: (1) The percentage of correctly streamed tiles. (2) The level of sDoP satisfaction.   
For arbitrary given predictor, sDoP, and resources, we consider the following privacy-aware QoE metric
\begin{align*}
	\mathrm{QoE} &\triangleq \frac{1}{L-l_0 + 1}\sum_{l=l_0}^L\underbrace{\frac{(\mathbf{q}_{l})^{\mathsf{T}}\cdot \mathbf{s}_{l}}{\|\mathbf{q}_{l}\|_1}}_{\textit{percent of correctly streamed tiles}}\cdot\underbrace{\frac{(\mathbf{q}_{l}^{\rho})^{\mathsf{T}}\cdot \mathbf{s}_{l}}{\|\mathbf{q}_{l}^{\rho}\|_1}}_{\textit{sDoP satisfaction}}
\end{align*}
where $\mathbf{s}_l\triangleq [s_{l,1},...,s_{l,M}]^\mathsf{T}$ denotes the selected tiles for streaming with  $s_{l,m}\in\{0,1\}$, $\mathbf{q}_{l}^{\rho}\triangleq [q_{l,1}^{\rho},...,q_{l,M}^{\rho}]^\mathsf{T}$ denotes the privacy-aware tile requests for the segment with $q_{l,m}^{\rho}\in\{0,1\}$. When the tiles are selected, $s_{l,m}=1$, otherwise $s_{l,m}=0$. When the tile is truly requested or camouflaged to be requested, $q_{l,m}^{\rho}=1$, otherwise $q_{l,m}^{\rho}=0$.

The number of selected tiles is limited by the CC capability, i.e., $\|\mathbf{s}_{l}\|_1 = C_{\mathrm{cc}}(t_{\mathrm{com}} ,t_{\mathrm{cpt}})\cdot M$. The number of privacy-aware requested tiles depends on sDoP, i.e., $\|\mathbf{q}_{l}^{\rho}\|_1 = N_{p}(\rho_s) $.
To gain useful insight, we assume that the selected tiles are the privacy-aware requested tiles, i.e., $\mathbf{s}_l = \mathbf{q}_l^{\rho}$. Hence the number of selected tiles and privacy-aware requested tiles are also identical, i.e.,
\begin{align}\label{constraint:sDoP_satasify}
	C_{\mathrm{cc}}(t_{\mathrm{com}} ,t_{\mathrm{cpt}})\cdot M = N_p(\rho_s)
\end{align}
Then, the privacy-aware QoE degenerates into
\begin{align}
\mathrm{QoE} &= \frac{1}{L-l_0 + 1}\sum_{l=l_0}^L\frac{(\mathbf{q}_{l})^{\mathsf{T}}\cdot \mathbf{q}_{l}^{\rho}}{\|\mathbf{q}_{l}\|_1}\label{qoe_cal}
\end{align}

We can observe that the QoE is affected by the average-DoO and the CC capability, which can be expressed as
\begin{align*}
	\mathrm{QoE} = \mathcal{Q}\left(\mathcal{D}\left(t_{\mathrm{obw}}\right), C_{\mathrm{cc}}(t_{\mathrm{com}}, t_{\mathrm{cpt}})\right)\in[0,100\%]\label{QoE-Metric}
\end{align*}
When the value of the QoE is $100\%$, all the truly requested tiles in a VR video are proactively computed and delivered before playback. Moreover, the MEC server only obtains the privacy-aware tile requests.

When $C_{\mathrm{cc}}(t_{\mathrm{com}}, t_{\mathrm{cpt}})$ is improved, more tiles can be rendered and transmitted, then more real requested tiles can be satisfied. When $\mathcal{D}\left(t_{\mathrm{obw}}\right)$ is improved, more of the streamed tiles are the real requested tiles. Then, we can find that the privacy-aware QoE monotonically increases with average-DoO and CC capability, respectively. 

\vspace{-0.3cm}
\section{sDoP: Contradictory roles for the QoE}
In this section, we investigate the role of sDoP for the privacy-aware QoE. To this end, we first optimize durations for observation window, communication, and computing to maximize the QoE. From the obtained closed-form solution, 
we investigate the impact of sDoP on average-DoO and CC capability, respectively. Finally, we discuss the overall impact of sDoP on the QoE.

\vspace{-4mm}
\subsection{Joint Optimization of the Durations for Prediction, Communication, and Computing}
Given arbitrary computing rate $C_{\mathrm{cpt}}$, transmission rate $C_{\mathrm{com}}$ and sDoP $\rho_s$, we aim to find the optimal durations for observation window, communication and computing to achieve the maximized QoE,
i.e.,
\begin{subequations}
	\begin{align}
		\textbf{P0}:&\ \ \ \  \ \max_{t_{\mathrm{obw}}, t_{\mathrm{cpt}},t_{\mathrm{com}}}  \mathcal{Q}\left(\mathcal{D}\left(t_{\mathrm{obw}}\right), C_{\mathrm{cc}}(t_{\mathrm{com}}, t_{\mathrm{cpt}})\right) \\
		&  \ \ \ \  \  s.t.  \ \   C_{\mathrm{cc}}(t_{\mathrm{com}}, t_{\mathrm{cpt}}) = \frac{N_{p}(\rho_s)}{M} \label{P0_C_cc}\\
		& \ \ \ \ \  \  \ \  \ \ \ t_{\mathrm{obw}} + t_{\mathrm{cpt}} + t_{\mathrm{com}} = T_{\mathrm{ps}} \label{P0_t_ps}
	\end{align}
\end{subequations}
As derived in the Appendix, the solution of \textbf{P0} is,
\begin{subequations}\label{P3:opt_solution_general}
	\begin{align}
		&t_{\mathrm{obw}}^{*}= \left\lfloor\left.  \left(T_{\mathrm{ps}} - \left(\frac{s_{\mathrm{com}}}{C_{\mathrm{com}}} + \frac{s_{\mathrm{cpt}}}{C_{\mathrm{cpt}}}\right)\cdot N_p(\rho_s)\right)\middle/\tau\right.\right\rfloor
		\label{P3:opt_solution_general_t_obw}\\
		&t_{\mathrm{com}}^{*}=\frac{s_{\mathrm{com}}N_p(\rho_s)}{C_{\mathrm{com}}} \ \ t_{\mathrm{cpt}}^{*}=\frac{s_{\mathrm{cpt}}N_p(\rho_s)}{C_{\mathrm{cpt}}}\label{P3:opt_solution_general_t_cc}
	\end{align}
\end{subequations}
where $\tau$ (in seconds) is the sampling interval of the user-behavior-related data in the observation window. The optimal duration for communication and computing $t_{\mathrm{cc}}^*$ can be obtained from \eqref{P3:opt_solution_general_t_cc} as
\begin{align}
	t_{\mathrm{cc}}^* = \left(\frac{s_{\mathrm{com}}}{C_{\mathrm{com}}} + \frac{s_{\mathrm{cpt}}}{C_{\mathrm{cpt}}}\right)\cdot N_p(\rho_s)\label{opt_t_cc}
\end{align}
By substituting \eqref{P3:opt_solution_general_t_cc} into \eqref{def:R_cc}, the maximized  resources rate is
\begin{align}
	R_{\mathrm{cc}}^* = \left. 1 \middle/ \left(\frac{s_{\mathrm{com}}M}{C_{\mathrm{com}}} + \frac{s_{\mathrm{cpt}}M}{C_{\mathrm{cpt}}}\right) \right.\label{opt_R_cc}
\end{align}
where $\left(\frac{s_{\mathrm{com}}M}{C_{\mathrm{com}}} + \frac{s_{\mathrm{cpt}}M}{C_{\mathrm{cpt}}}\right)$ is the  required optimal duration to render and transmit all tiles in a segment. 
By substituting \eqref{opt_R_cc} into  \eqref{P3:opt_solution_general_t_obw}, $t_{\mathrm{obw}}^*$ can also be expressed as
\begin{align}
	t_{\mathrm{obw}}^* = \left\lfloor\left.  \left(T_{\mathrm{ps}} - \frac{N_{p}(\rho_s)}{R_{\mathrm{cc}}^* \cdot M}\right)\middle/\tau\right.\right\rfloor\label{t_obw_cal}
\end{align}


%

\subsection{Contradictory Roles of sDoP}

\subsubsection{Improve the CC Capability}
By substituting \eqref{opt_t_cc} and \eqref{opt_R_cc} into \eqref{def:R_cc}, the CC capability can be rewritten as follows:
\begin{align*}
	C_{\mathrm{cc}}(t_{\mathrm{com}}^*, t_{\mathrm{cpt}}^*) = R_{\mathrm{cc}}^* t_{\mathrm{cc}}^*,
\end{align*}
Further considering \eqref{opt_t_cc} and \eqref{opt_R_cc}, we can find that the resources rate $R_{\mathrm{cc}}^*$ has no relation with $\rho_s$. Besides, the optimal total duration for communication and computing $t_{\mathrm{cc}}^*$ increases with $\rho_s$. This means that the increase of $\rho_s$ improves the CC capability. 

\subsubsection{Degrade the Average-DoO}
With the increase of $\rho_s$, the duration of observation window $t_{\mathrm{obw}}^* = T_{\mathrm{ps}} - t_{\mathrm{cc}}^*$ will be reduced, which can also be verified from \eqref{P3:opt_solution_general_t_obw}. The reduction of $t_{\mathrm{obw}}^*$ degrades the average-DoO.

\begin{figure}[htbp]
	\vspace{-0.5cm}
	\centering
	\begin{minipage}[t]{0.85\linewidth}
		\includegraphics[width=1\textwidth]{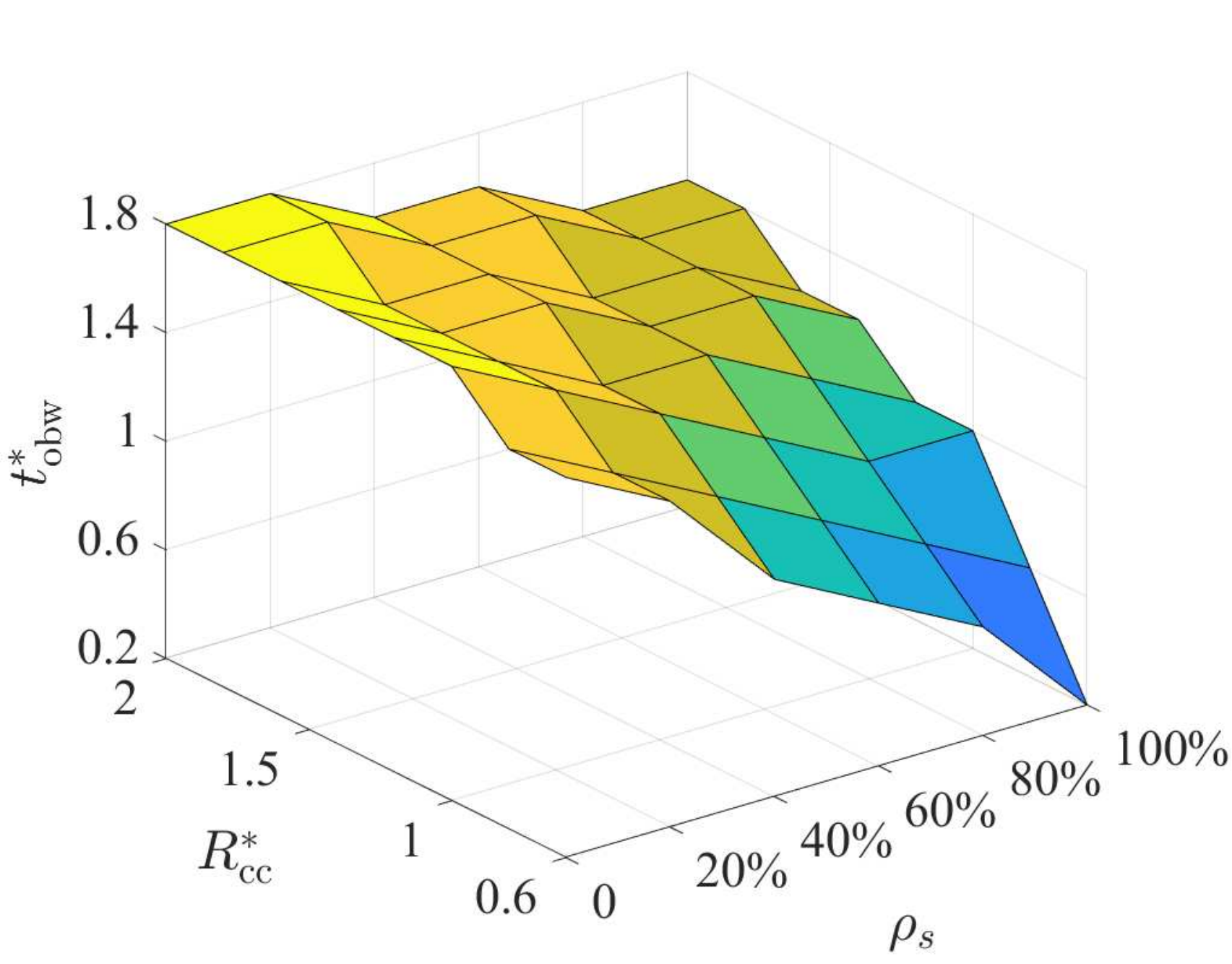}
	\end{minipage}
	\caption{$t_{\mathrm{obw}}^*$ v.s. $R_{\mathrm{cc}}^*$ and $\rho_s$.}
	\label{Fig:Resource-saturated_Region}
	\vspace{-0.5cm}
\end{figure}

In Fig. \ref{Fig:Resource-saturated_Region}, we use the values of $t_{\mathrm{obw}}^*$ obtained from \eqref{P3:opt_solution_general_t_obw} to visualize the impact of $\rho_s$ on the average-DoO. 
We can observe that as the increase of $\rho_s$, the reduction of $t_{\mathrm{obw}}^*$ is discrete, which comes from the discrete sampling in the observation window. Besides, the value of $t_{\mathrm{obw}}^*$ increases with the maximized resources rate $R_{\mathrm{cc}}^*$. This is because when the resources rate is increased, privacy-aware requested tiles can be streamed with less duration, i.e., $t_{\mathrm{cc}}^*$ is reduced. Then, $t_{\mathrm{obw}}^* = T_{\mathrm{ps}} - t_{\mathrm{cc}}^*$ is increased.

\subsubsection{Overall Impact} 
For the final QoE, the impact of sDoP is complicated. On the one hand, it improves the CC capability. On the other hand, it degrades the average-DoO. Generality speaking, the overall effect depends on if the QoE is dominated by the increase of CC capability or the reduction of the average-DoO.  

\vspace{-0.4cm}
\section{Trace-Driven Simulation Results}
\begin{table*}[htbp]
	\captionsetup{font={small}}
	\vspace{-0.5cm}
	\caption{Settings of VR video}\label{table:LR}
	\vspace{-0.4cm}
	\begin{center}
		\begin{tabular}{|c|c|c|c|}
			\hline
			Resolution & 3840$\times$2160 pixels\cite{FoV_aware_tile} &$b$ &$12$ bits per pixel\cite{HuaWei_Cloud_VR}\\
			\hline 
			Number of tiles $M$ & 10 rows $\times$ 20 columns = 200 & Frame rate $r_f$ & 30 FPS \cite{FoV_aware_tile} \\ \hline
			Pixels in wide of a tile ${px}_w$ & $3840/20 = 192$&Pixels in height of a tile ${px}_h$ & $2160/10=216$\\ \hline
			Compression ratio $\gamma_c$ & 2.41\cite{HEVC_lossless_coding}& Playback duration of a segment $T_{\mathrm{seg}}$ & 1 s \cite{FoV_aware_tile}\\ \hline
			Number of bits in a tile for transmission $s_{\mathrm{com}}$ & 5.9 Mbits$^a$ & Number of bits in a tile for rendering $s_{\mathrm{cpt}}$ & 14.2 Mbits$^a$\\ \hline
			Size of FoV & $100^{\circ}\times100^{\circ}$ circles \cite{FoV_size_NTHU,Romero_ACM_MMsys_20_code} & Number of tiles in an FoV $N_{\textit{fov}}$ & 33$^b$\\ \hline
		\end{tabular}
		\footnotesize{ $^a$These values are calculated from the definition in Section \ref{section:s_com,s_cpt}. $^b$This is the average value on the dataset \cite{NTHU_dataset}, according to our tests. }
	\end{center}
	\vspace{-0.6cm}
\end{table*}

In this section, we show the overall impact of sDoP on QoE via trace-driven simulation results. 
First, we consider the prediction task on a real dataset \cite{NTHU_dataset}, where 300 traces of head movement positions from 30 users watching 10 VR videos are used for training and testing predictors.\footnote{According to the analysis in  \cite{Romero_ACM_MMsys_20_code,Romero_ACM_MMsys_20_paper}, the traces of the first 20 users in the dataset have mistakes, thus we only use the traces of the other 30 users.} We randomly split the total traces into training and testing sets with the ratio 8:2.

We use two predictors, \textit{position-only} and \textit{no-motion} predictors, which achieve the state-of-the-art accuracy for the dataset, according to tests in \cite{Romero_ACM_MMsys_20_code}. The \textit{position-only} predictor employs a sequence-to-sequence LSTM-based architecture, which uses the time series of past head movement positions as input, to predict the time series of future positions \cite{Romero_ACM_MMsys_20_code}. The predictor does not consider the time required for computing and communication as well as the spatial degree of privacy. To reserve time for computing and communication, we tailor the predictor as follows. Set the duration between the end of the observation window and the beginning of the prediction windows as $t_{\mathrm{cc}}^*$, set the durations of observation and prediction windows as $t_{\mathrm{obw}}^*$ and $T_{\mathrm{seg}}$, respectively. To satisfy the privacy requirement, we consider a classical federated learning, \texttt{FederatedAveraging} algorithm in \cite{google_federated_learning_17}. The settings of the federated learning are as follows. For each round, we select \textit{all} of $K=30$ users to update the model parameters of the predictor. The number of local epochs for each user is $E_l=50$, the number of communication rounds is $R=10$. Hence, every trace in the train set is used $50\times10=500$ times for training, which is consistent with the centralized training \cite{Romero_ACM_MMsys_20_code}. The weighting coefficient of the $k$th user on the model parameter is $c_k =\frac{n_k}{N_{\textit{train}}}$, where $N_{\textit{train}}=300\times0.8=240$ is the total number of traces in the train set, $n_k$ is the number of video traces that belong to the $k$th user in the training set. Due to the random division of training and testing sets, $n_k$ varies from 6 to 10.
We refer to the predictor as \textit{tailored federated position-only} predictor. Other details and hyper-parameters of the tailored predictor are the same as the \textit{position-only} predictor \cite{Romero_ACM_MMsys_20_code}. The \textit{no-motion} predictor simply uses the last position in the observation window as the predicted time series of future positions \cite{Romero_ACM_MMsys_20_code}.

The maximized resources rate $R_{\mathrm{cc}}^*$ depends on the configured communication and computing resources as well as the number of users. For example, when $K=4$, $N_t=8$, $P=24$ dBm, $B=150$ MHz, and $d_k=5$ m, the ensemble-average transmission rate for a user is $C_{\mathrm{com}}=2.85$ Gbps\cite{Xing_VR_Shannon}. When Nvidia RTX 8000 GPU is used for rendering VR videos for four users, the computing rate for a user is $C_{\mathrm{cpt}}=2.2$ Gbps \cite{Xing_VR_Shannon}. Then, the maximized resources rate is $R_{\mathrm{cc}}^* = \left. 1 \middle/ \left(\frac{s_{\mathrm{com}}M}{C_{\mathrm{com}}} + \frac{s_{\mathrm{cpt}}M}{C_{\mathrm{cpt}}}\right) \right. = 0.6$. To reflect the variation of configured resources, we set $R_{\mathrm{cc}}^*\in[0.6,2]$.

The settings of VR video are listed in Table \ref{table:LR}.
To gain useful insight, we assume that all users have identical sDoP requirement among all videos, ranging from 0 to 100\%. The procedure of simulation is given in \textbf{Procedure \ref{alg:Simulation_Procedure}}.

\renewcommand{\algorithmicrequire}{ \textbf{Input:}} 
\renewcommand{\algorithmicensure}{ \textbf{Output:}} 
\floatname{algorithm}{ Procedure}
\begin{algorithm}[htb]
	\caption{Obtain the average QoE, $\mathcal{D}(t_{\mathrm{obw}}^*)$, and $C_{\mathrm{cc}}(t_{\mathrm{com}}^*, t_{\mathrm{cpt}}^*)$}
	\label{alg:Simulation_Procedure}
	\begin{algorithmic}
		\REQUIRE set $\rho_s\in[0,100\%]$, $R_{\mathrm{cc}}^*\in[0.6,2]$.
		\STATE \textbf{(i)}. Given sDoP $\rho_s$ and maximized resource rates $R_{\mathrm{cc}}^*$, obtain the values of the optimal duration of observation window $t_{\mathrm{obw}}^*$ from \eqref{t_obw_cal}.
		\STATE \textbf{(ii)}. Train the \textit{tailored federated position-only} predictor with different $t_{\mathrm{obw}}^*$.
		\STATE \textbf{(iii)}. Given $t_{\mathrm{obw}}^*$ and two predictors, obtain the predicted time series of positions.
		\STATE \textbf{(iv)}. With $N_{\textit{fov}}$, determine the predicted requested tiles by scheme (b) of Section IV-C in \cite{privacy_Xing}.
		\STATE \textbf{(v)}. Given sDoP, obtain the CC capability from \eqref{constraint:sDoP_satasify}. 
		\STATE \textbf{(vi)}. Given the CC capability, the predicted requested tiles, and masking scheme (a) in Fig. \ref{Fig:sDoP_explain}, determine the extra camouflaged requested tiles.
		\STATE \textbf{(vii)}. Calculate the QoE from \eqref{qoe_cal}, then average over the testing set.
		
	\end{algorithmic}
\end{algorithm}

\begin{figure}[htbp]
	\vspace{-0.2cm}
	\centering
	\subfloat[No-motion predictor]{\label{Fig:qoe_rho_T_cc_m_No_motion}
		\begin{minipage}[c]{0.85\linewidth}
			\centering
			\includegraphics[width=1\textwidth]{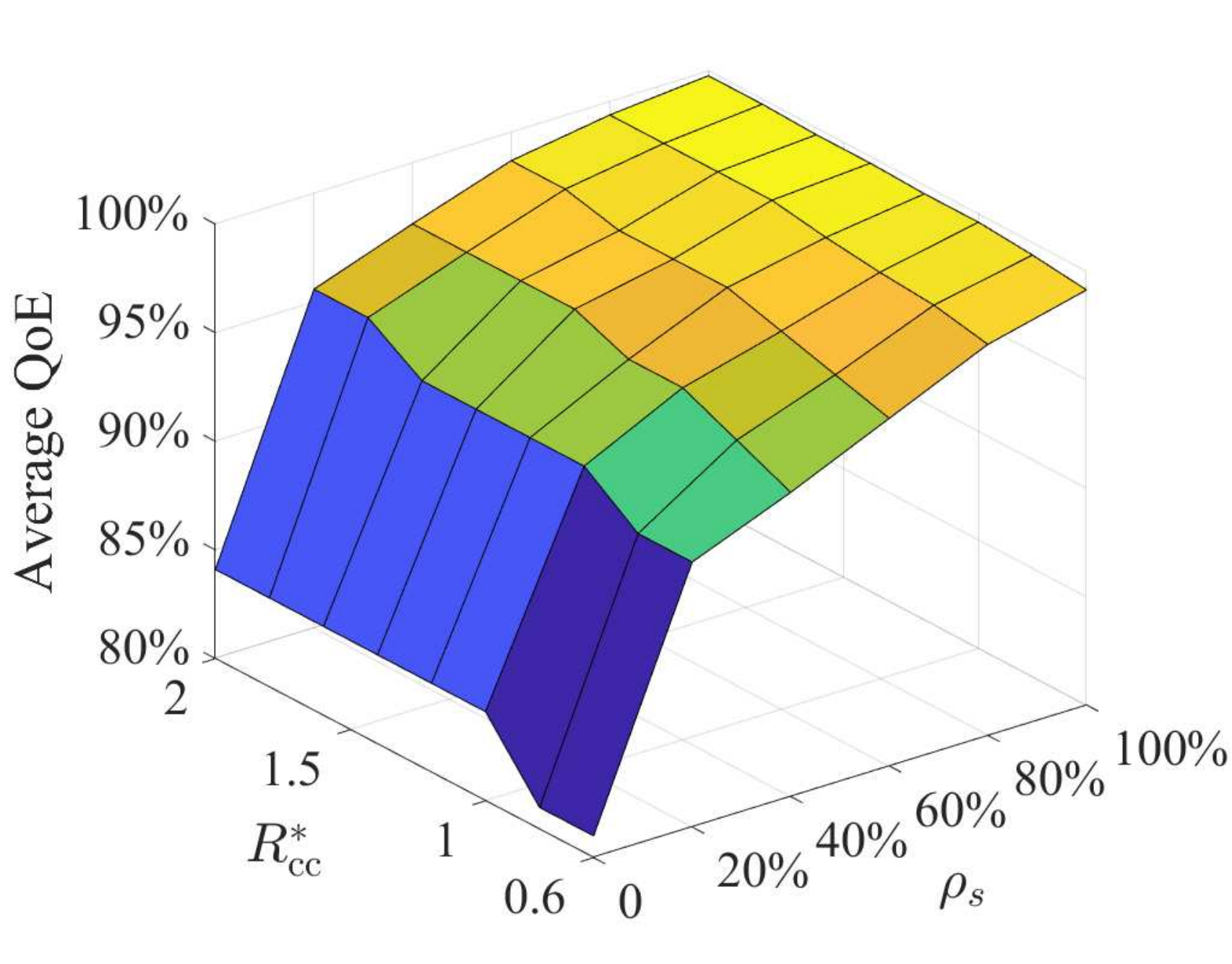}
		\end{minipage}
	}\\ \vspace{-0.4cm}
	\subfloat[Tailored federated postion-only predictor]{\label{Fig:qoe_rho_T_cc_m_pos_only}
		\begin{minipage}[c]{0.85\linewidth}
			\centering
			\includegraphics[width=1\textwidth]{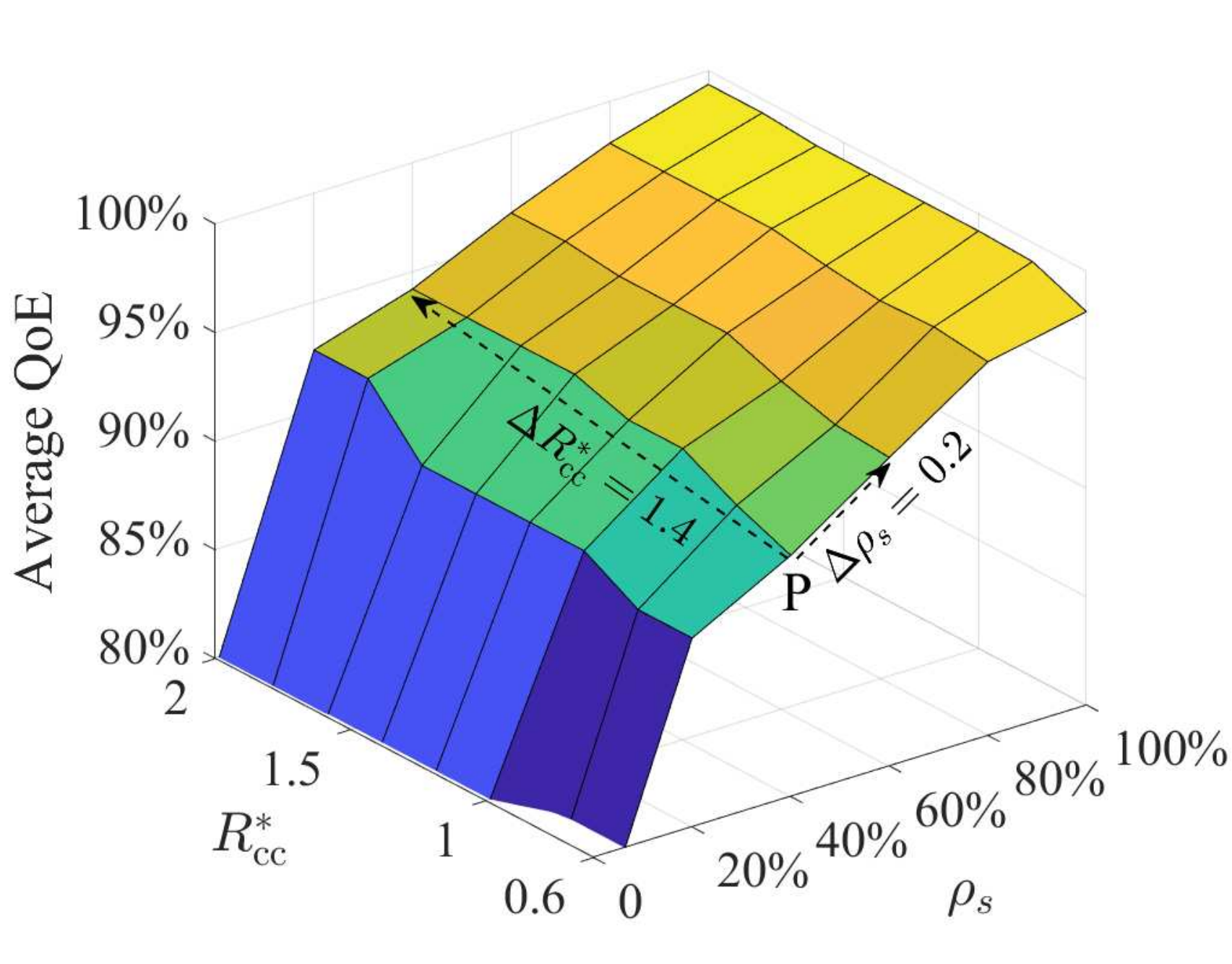}
		\end{minipage}
	}\vspace{-0.2cm}
	\caption{Average QoE v.s. resource rates and sDoP.}\label{Fig:qoe_rho_T_cc_m}
	\vspace{-0.4cm}
\end{figure}

In Fig. \ref{Fig:qoe_rho_T_cc_m}, we show the average QoE achieved by two predictors versus the assigned resources rate and sDoP. We can observe that no matter how much the resources rate is assigned, which predictor is employed, the average QoE can always be improved with the increase of $\rho_s$. Besides, \textbf{the increase of $\rho_s$ is equivalent to the increase of assigned resources rate in terms of improving the QoE}. For example, consider the point ``P" in Fig. \ref{Fig:qoe_rho_T_cc_m_pos_only}. To achieve QoE = 94\%, increasing $\rho_s$ by 0.2 is equivalent to increasing $R_{\mathrm{cc}}^*$ by 1.4. 

To further understand how the QoE is affected by the sDoP, we consider a case where the resource rates $R_{\mathrm{cc}}^*=0.6$ as an example, to investigate how the CC capability $C_{\mathrm{cc}}(t_{\mathrm{com}}^*, t_{\mathrm{cpt}}^*)$, average-DoO $\mathcal{D}(t_{\mathrm{obw}}^*)$, and average QoE is affected by $\rho_s$. As shown in Fig. \ref{Fig:QoE, CC capability, and DoO v.s. DoP}, for both predictors, the increase of sDoP improves the CC capability $C_{\mathrm{cc}}(t_{\mathrm{com}}^*, t_{\mathrm{cpt}}^*)$ and degrades average-DoO $\mathcal{D}(t_{\mathrm{obw}}^*)$. Since the degradation of average-DoO is relative small, the QoE is dominated by the increase of the CC capability. 

\begin{figure}[htbp]
	\vspace{-0.4cm}
	\centering
	\subfloat[No-motion predictor]{\label{Fig:qoe_rho_C_cc_No_motion_T_cc_m_6}
		\begin{minipage}[c]{0.85\linewidth}
			\centering
			\includegraphics[width=1\textwidth]{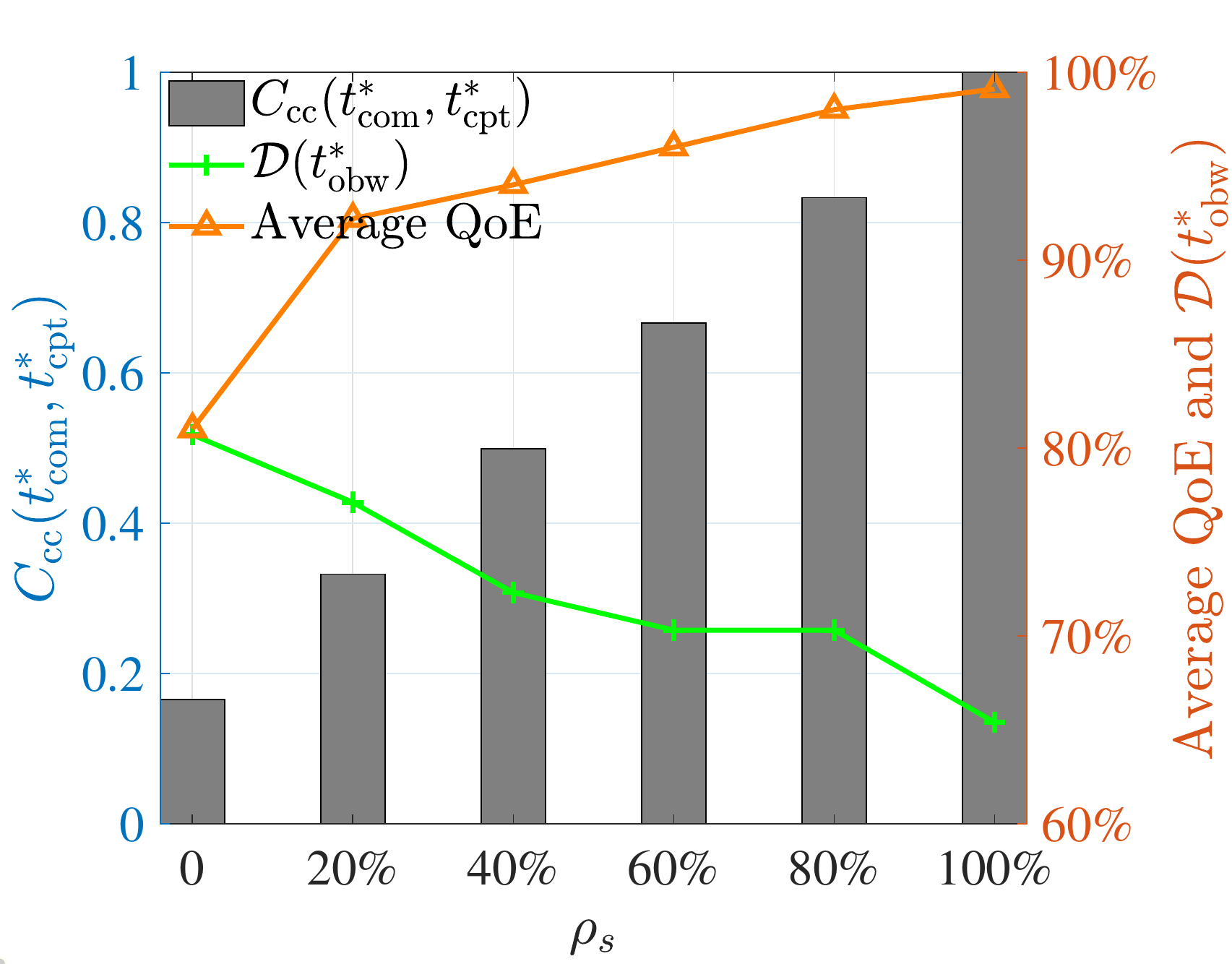}
		\end{minipage}
	}\\ \vspace{-0.4cm}
	\subfloat[Tailored federated position-only predictor]{\label{Fig:qoe_rho_C_cc_Pos_Only_T_cc_m_6}
		\begin{minipage}[c]{0.85\linewidth}
			\centering
			\includegraphics[width=1\textwidth]{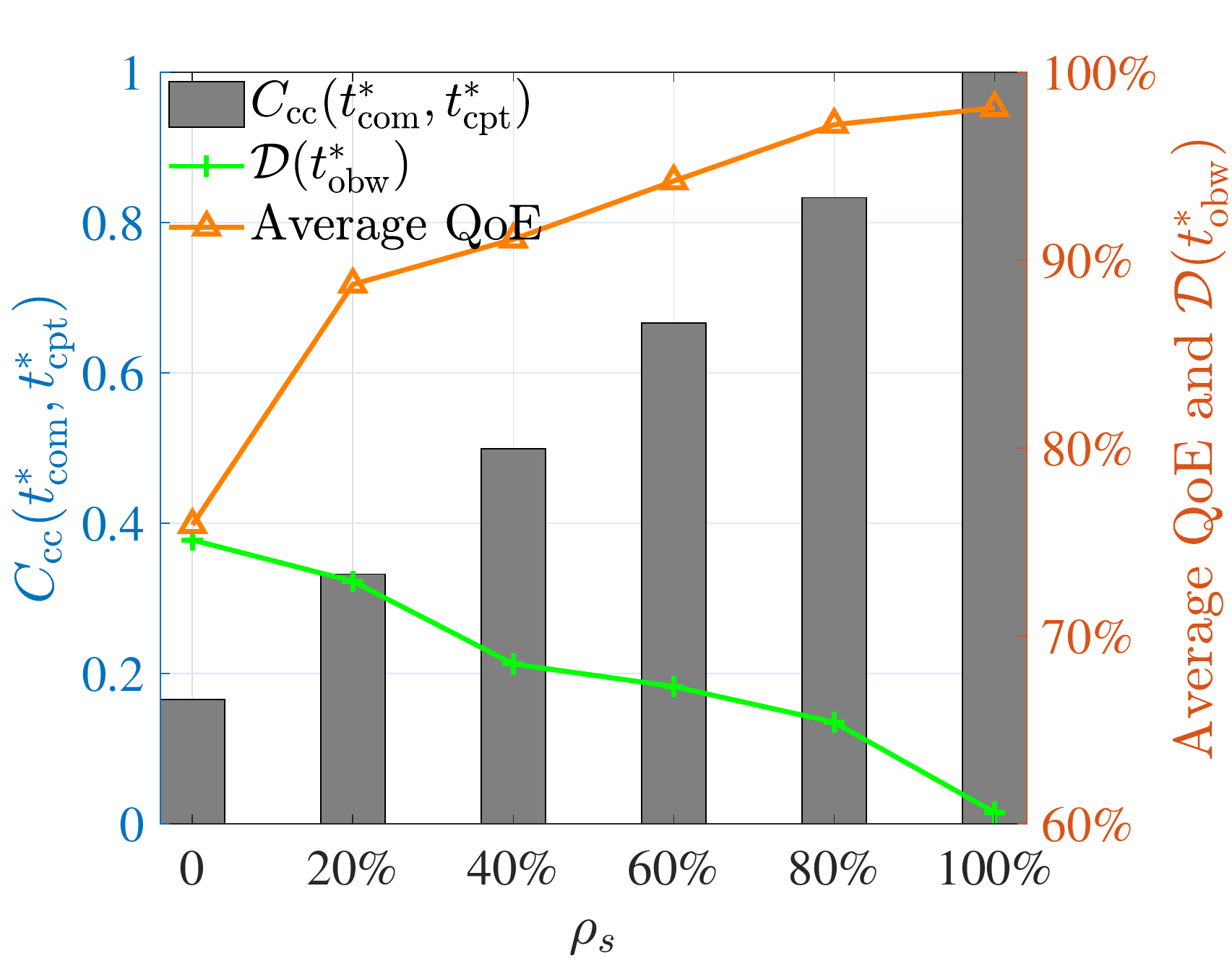}
		\end{minipage}
	}\vspace{-0.2cm}
	\caption{Average QoE, CC capability, and average-DoO v.s. sDoP, $R_{\mathrm{cc}}^*=0.6$.}\label{Fig:QoE, CC capability, and DoO v.s. DoP}
	\vspace{-0.6cm}	
\end{figure}

\vspace{-0.15cm}
\section{Conclusion}
\vspace{-0.15cm}
In this paper, we defined spatial privacy requirement for better privacy protection and investigated the impact of spatial privacy requirement on VR video streaming. By duration optimization and analyzing the obtained optimal closed-form solution, we found the relation between sDoP and QoE. The analysis showed that the increase of sDoP improves the CC capability but degrades the average-DoO. The overall impact of sDoP on QoE depends on which factor dominates the QoE. Simulation with two predictors on a real dataset validated the analysis and showed that the overall impact of sDoP is to improve the QoE.

\bibliographystyle{IEEEtran}
\bibliography{IEEEabrv,ref}

\begin{appendices}
	\section{Proof of the Solution of Problem \textbf{P0}}
	We can observe that for arbitrary $\rho_s$, CC capability is fixed as $C_{\mathrm{cc}}(t_{\mathrm{com}},t_{\mathrm{cpt}}) = \frac{N_{p}(\rho_s)}{M}$. Then, QoE becomes a function of a single variable $\mathcal{D}(t_{\mathrm{obw}})$. Furthering consider \textbf{Remark 1}, 
	we can find maximizing $\mathcal{D}(t_{\mathrm{obw}})$ is equivalent to maximize $t_{\mathrm{obw}}=T_{\mathrm{ps}} - (t_{\mathrm{com}} + t_{\mathrm{cpt}})$. Then, without loss of optimally, \textbf{P0} can be transformed as 
	\begin{subequations}\label{P1}
		\begin{align}
			&\ \ \ \  \ \min_{t_{\mathrm{cpt}},t_{\mathrm{com}}}  t_{\mathrm{com}} + t_{\mathrm{cpt}} \\
			&  \ \ \ \  \  s.t.  \ \   \min\left\{\frac{C_{\mathrm{com}}t_{\mathrm{com}}}{s_{\mathrm{com}}^{}}, \frac{C_{\mathrm{cpt}}t_{\mathrm{cpt}}}{s_{\mathrm{cpt}}^{}}, M\right\} = N_{p}(\rho_s) \label{P1_C_N},
		\end{align}
	\end{subequations}
	where \eqref{P1_C_N} can be obtained by multiplying both sides of \eqref{P0_C_cc} by $M$. Since $N_{p}(\rho_s)\leq M$, \eqref{P1_C_N} can be further simplified as $\left\{\frac{C_{\mathrm{com}}t_{\mathrm{com}}}{s_{\mathrm{com}}^{}}, \frac{C_{\mathrm{cpt}}t_{\mathrm{cpt}}}{s_{\mathrm{cpt}}^{}}\right\} = N_{p}(\rho_s)$. According to the constraint, we discuss problem \eqref{P1} in the following three cases.
	
	(1) When $\frac{C_{\mathrm{com}}t_{\mathrm{com}}}{s_{\mathrm{com}}^{}}> \frac{C_{\mathrm{cpt}}t_{\mathrm{cpt}}}{s_{\mathrm{cpt}}^{}}$, upon substituting into \eqref{P1_C_N}, we have $t_{\mathrm{cpt}}=\frac{N_p(\rho_s)s_{\mathrm{cpt}}}{C_{\mathrm{cpt}}}$. Upon substituting into the condition $\frac{C_{\mathrm{com}}t_{\mathrm{com}}}{s_{\mathrm{com}}^{}}> \frac{C_{\mathrm{cpt}}t_{\mathrm{cpt}}}{s_{\mathrm{cpt}}^{}}$, we obtain $t_{\mathrm{com}}\geq\frac{N_p(\rho_s)s_{\mathrm{com}}}{C_{\mathrm{com}}}$. Then, we obtain $t_{\mathrm{com}} + t_{\mathrm{cpt}}>\frac{s_{\mathrm{com}}N_p(\rho_s)}{C_{\mathrm{com}}} + \frac{s_{\mathrm{cpt}}N_p(\rho_s)}{C_{\mathrm{cpt}}}$.
	
	(2) When $\frac{C_{\mathrm{com}}t_{\mathrm{com}}}{s_{\mathrm{com}}^{}}< \frac{C_{\mathrm{cpt}}t_{\mathrm{cpt}}}{s_{\mathrm{cpt}}^{}}$, upon substituting into \eqref{P1_C_N}, we have $t_{\mathrm{com}}=\frac{N_p(\rho_s)s_{\mathrm{com}}}{C_{\mathrm{com}}}$. Upon substituting the condition $\frac{C_{\mathrm{com}}t_{\mathrm{com}}}{s_{\mathrm{com}}^{}}< \frac{C_{\mathrm{cpt}}t_{\mathrm{cpt}}}{s_{\mathrm{cpt}}^{}}$, we obtain $t_{\mathrm{cpt}}>\frac{N_p(\rho_s)s_{\mathrm{cpt}}}{C_{\mathrm{cpt}}}$. Then, we obtain $t_{\mathrm{com}} + t_{\mathrm{cpt}}>\frac{s_{\mathrm{com}}N_p(\rho_s)}{C_{\mathrm{com}}} + \frac{s_{\mathrm{cpt}}N_p(\rho_s)}{C_{\mathrm{cpt}}}$.
	
	(3) When $\frac{C_{\mathrm{com}}t_{\mathrm{com}}}{s_{\mathrm{com}}^{}} = \frac{C_{\mathrm{cpt}}t_{\mathrm{cpt}}}{s_{\mathrm{cpt}}^{}}$, upon substituting into \eqref{P1_C_N}, we have $t_{\mathrm{com}}=\frac{N_p(\rho_s)s_{\mathrm{com}}}{C_{\mathrm{com}}}$ and $t_{\mathrm{cpt}}=\frac{N_p(\rho_s)s_{\mathrm{cpt}}}{C_{\mathrm{cpt}}}$. Then, we obtain $t_{\mathrm{com}} + t_{\mathrm{cpt}}=\frac{s_{\mathrm{com}}N_p(\rho_s)}{C_{\mathrm{com}}} + \frac{s_{\mathrm{cpt}}N_p(\rho_s)}{C_{\mathrm{cpt}}}$.
	
	From the three cases, we obtain the solution as
	$t_{\mathrm{com}}^{*}=\frac{s_{\mathrm{com}}N_p(\rho_s)}{C_{\mathrm{com}}}$, $t_{\mathrm{cpt}}^{*}=\frac{s_{\mathrm{cpt}}N_p(\rho_s)}{C_{\mathrm{cpt}}}$.
	Upon substituting into \eqref{P0_t_ps}, $t_{\mathrm{obw}}$ can be obtained as $t_{\mathrm{obw}}=T_{\mathrm{ps}} - \left(\frac{s_{\mathrm{com}}}{C_{\mathrm{com}}} + \frac{s_{\mathrm{cpt}}}{C_{\mathrm{cpt}}}\right)\cdot N_p(\rho_s)$.
	Note that although the duration of the observation window can be continuous, the user behavior-related data is sampled discretely. 
	Then, the efficient duration of the observation window is
	\begin{align}
		t_{\mathrm{obw}}^{*}= \left\lfloor\left.  \left(T_{\mathrm{ps}} - \left(\frac{s_{\mathrm{com}}}{C_{\mathrm{com}}} + \frac{s_{\mathrm{cpt}}}{C_{\mathrm{cpt}}}\right)\cdot N_p(\rho_s)\right)\middle/\tau\right.\right\rfloor
	\end{align}
\end{appendices}

\end{document}